\documentclass[review]{elsarticle}
\makeatletter
\def\ps@pprintTitle{
	\let\@oddhead\@empty
	\let\@evenhead\@empty
	\def\@oddfoot{\reset@font\hfil\thepage\hfil}
	\let\@evenfoot\@oddfoot
}
\makeatother

 \biboptions{comma,sort&compress}

\usepackage[table]{xcolor}
\usepackage{color}
\usepackage{graphicx}
\usepackage{amsmath}
\usepackage{here}
\usepackage{cuted}

\usepackage{here}
\usepackage[dvips]{epsfig}
\definecolor{bluette}{rgb}{.2,.4,0}
\definecolor{salmon}{rgb}{.9,0.68,0.5}
\definecolor{motive}{rgb}{0.2,1,.5}
\definecolor{list}{rgb}{0.3,.8,.1}
\definecolor{moe}{rgb}{1,.7,.5}
\definecolor{mote}{rgb}{.7,.5,.6}
\definecolor{pisello}{rgb}{.1,1,0}
\definecolor{orange}{rgb}{1,.7,0}
\definecolor{oliva}{rgb}{.1,.5,0.3}
\definecolor{greenda}{rgb}{0,.3,.2}
\definecolor{greenli}{rgb}{0.2,.8,.0}
\definecolor{blueda}{rgb}{0,.1,.6}
\definecolor{purple}{rgb}{.7,.1,.2}
\definecolor{marrone}{rgb}{1,0.7,0}
\definecolor{pinky}{rgb}{1,0.8,0.8}
\definecolor{rose}{rgb}{1,0.4,0}
\definecolor{oro}{rgb}{0.6,.4,.0}

\def\oliva{\color{oliva}}

\def\purple{\color{purple}}
\def\blueda{\color{blueda}}
\def\blueli{\color{bluette}}

\def\beq{\begin{equation}}
\def\eeq{\end{equation}}
\def\bea{\begin{eqnarray}}
\def\eea{\end{eqnarray}}
\def\bq{\begin{quote}}
\def\eq{\end{quote}}

\def\nnb{\nonumber}
\def\ga{\left(}
\def\dr{\right)}

\def\rar{\rightarrow}

\def\nnb{\nonumber}

\def\la{\langle}
\def\ra{\rangle}

\def\ba{\vspace*{-0.2cm}\begin{array}}
\def\ea{\end{array}\vspace*{-0.2cm}}

\def\b{$\bullet~$}
\def\d{$\diamond~$}

\def\als{\alpha_s}

\def\gg2{\la\alpha_s G^2 \ra}
\def\gg3{g^3f_{abc}\la G^aG^bG^c \ra}
\def\ggg4{\la\als^2G^4\ra}

\def\gg{\lag g^{2}_{s} G^2 \rag}
\def\ggg{\lag g^{3}_{s}G^3\rag}

\usepackage{amsmath}
\usepackage{slashed}
\usepackage{color}

\begin{document}

\markboth{Stephan Narison, Montpellier (FR)}{ }
\begin{frontmatter}

\title{ 
 Muon and Tau Anomalies and $\Delta \alpha_{had}^{(5)} (M_Z^2)$ in the Standard Model\,\tnoteref{invit}}
\tnotetext[invit]{Talk given at QCD26 (20-24 july 2026, Montpellier-FR) and at HEPMAD26 (26-31 october 2026, Ambohidratrimo - Antananarivo - MG) based on the series of recent papers\,\cite{SN1,SN2,SN3,SN4,SN5} and is a re-make of the publication of my 3rd cycle thesis at the CPT of Marseille 50 years ago\,\cite{CALMET,SN78} and of the paper in Ref.\,\cite{SN01}. 
This review is a Tribute and Recognition to my former supevisor Professor Eduardo de Rafael being one of the leading Figure in this field who left us on 19th August 2026. He has been inivited to givea similar talk at QCD26.}
\author{Stephan Narison
}
\address{Laboratoire
Univers et Particules de Montpellier (LUPM), CNRS-IN2P3 and Univ. Montpellier, \\
Case 070, Place Eug\`ene
Bataillon, 34095 - Montpellier, France\\
and\\
Institute of High-Energy Physics of Madagascar (iHEPMAD)\\
University of Antananarivo, Ankatso 101, Madagascar}
\ead{snarison@yahoo.fr}


\date{\today}
\begin{abstract}

I review the present status of the Standard Model (SM) determination of the muon anomaly $a_\mu=\frac{1}{2}(g_\mu-2)$ by emphasizing the estimate of the lowest order (LO) hadronic vacuum polarization (HVP) contributions using recent $e^+e^-$ data on $2\pi$\,\cite{CMD3} and $3\pi$\,\cite{BABAR,BELLE}, the inclusive ratio $R^{ee}\equiv \sigma(e^+e^-\to$ Hadrons)/$\sigma (e^+e^-\to \mu^+\mu^-)$ of total  cross-sections compiled by PDG\,\cite{PDG}  to which we add the QCD continuum including higher order (HO) PT QCD up to order $\alpha_s^5$, quark mass corrections $m_q^2$ up to $\alpha_s^3$, $m_s^4$ up to $\alpha_s^2$ and non-pertubative QCD condensates. Adding the $e^+e^-\to \pi^0\gamma + \eta\gamma$ contributions to the CMD3 data, one obtains\,: $a_\mu\vert^{hvp}_{lo}=(7094\pm 37)\times 10^{-11}$. Combining this LO  result with the existing HO estimate of the hadronic, QED and EW contributions, one obtains\,: $\Delta a_\mu^{sm} \equiv (a_\mu^{exp}-a_\mu^{sm})= (86\pm 38_{th}\pm 14.5_{exp})\times 10^{-11} $. The average of the recent estimates quoted in Table\,3 [$\oplus$ attempt to fit all $\,e^+e^-$ data\,\cite{BZ26}] gives\,: $\Delta a_\mu^{sm} = (84.9[90.0]\pm 21.3_{th}\pm 14.5_{exp})\times 10^{-11}$\,.  These results agree with the precise BMW20 lattice one\,\cite{BMW} 
 and still suggest a window for new physics beyond the SM  of about $(2.1\sim 3.5)\,\sigma$. Besides some future projects for improving the $e^+e^-$ data, the MUonE for extracting directly $a_\mu^{tot}\vert^{hvp}_{lo}$ and J-PARC (E34) for an independent measurement of $a_\mu$, it is also extremely important to improve the Lattice results shown in Fig.\,9 which (at first sight) disagree each others despite the large errors and may question the reliability of the {\it averages of different averages} used in the White Paper (WP25)\,\cite{WP25}. Our analysis is completed by updating the estimate of the hadronic contribution to the tau anomaly\,: $a_\tau\vert^{hvp}_{lo}=(3516\pm 25)\times 10^{-9}$ and to $\alpha(M_Z)\,: \,\Delta \alpha_{had}^{(5)} (M_Z^2)= (2770.7\pm 4.5)\times 10^{-5}. $

\end{abstract}
\end{frontmatter}
\newpage
\section{Introduction to the lepton anomaly}
\section*{
\b Definition of the lepton anomaly}
\d {The magnetic moment of a lepton having a mass $m$ measures the force and the orientation of the magnetic field produced by the electric charge $e$ and of its spin  $\vec{S}\equiv 1/2$ (intrinsic angular momentum which differs with the orbital angular momentum)\,: 
\beq
 \vec{\mu}_S ={g}\frac{e}{2m_l} \vec{S}~.
\eeq
where ${ g}$ is the Land\'e \,Factor\, (1921).
\begin{figure}[hbt]
\begin{center}
\includegraphics[width=2.cm]{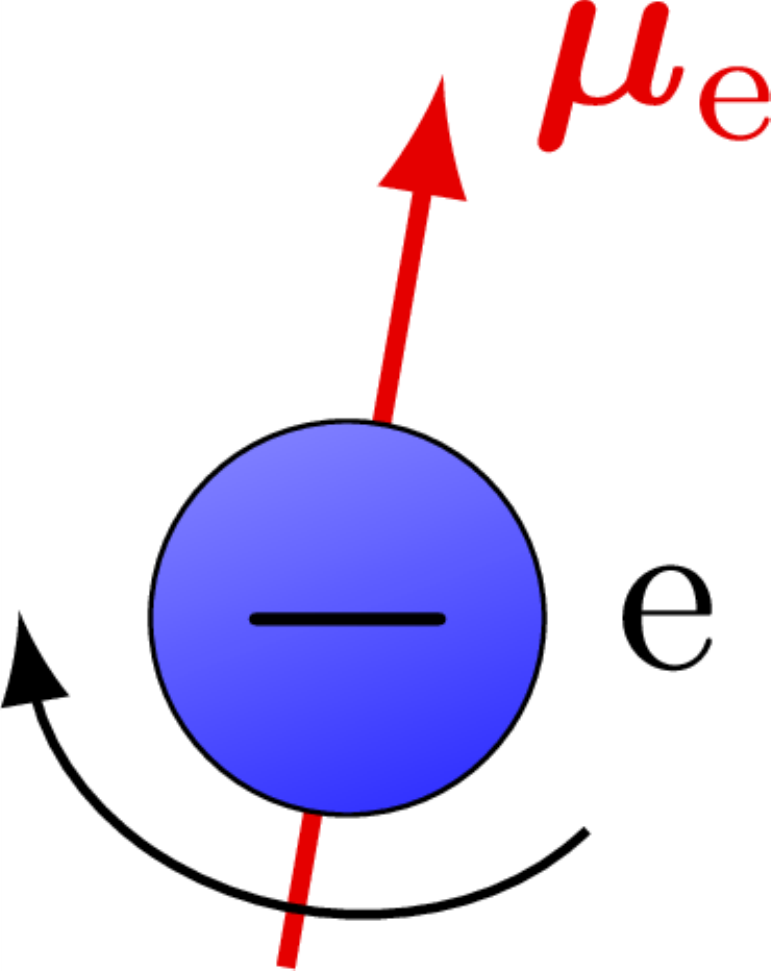}
\hspace*{2cm}\includegraphics[width=2.cm]{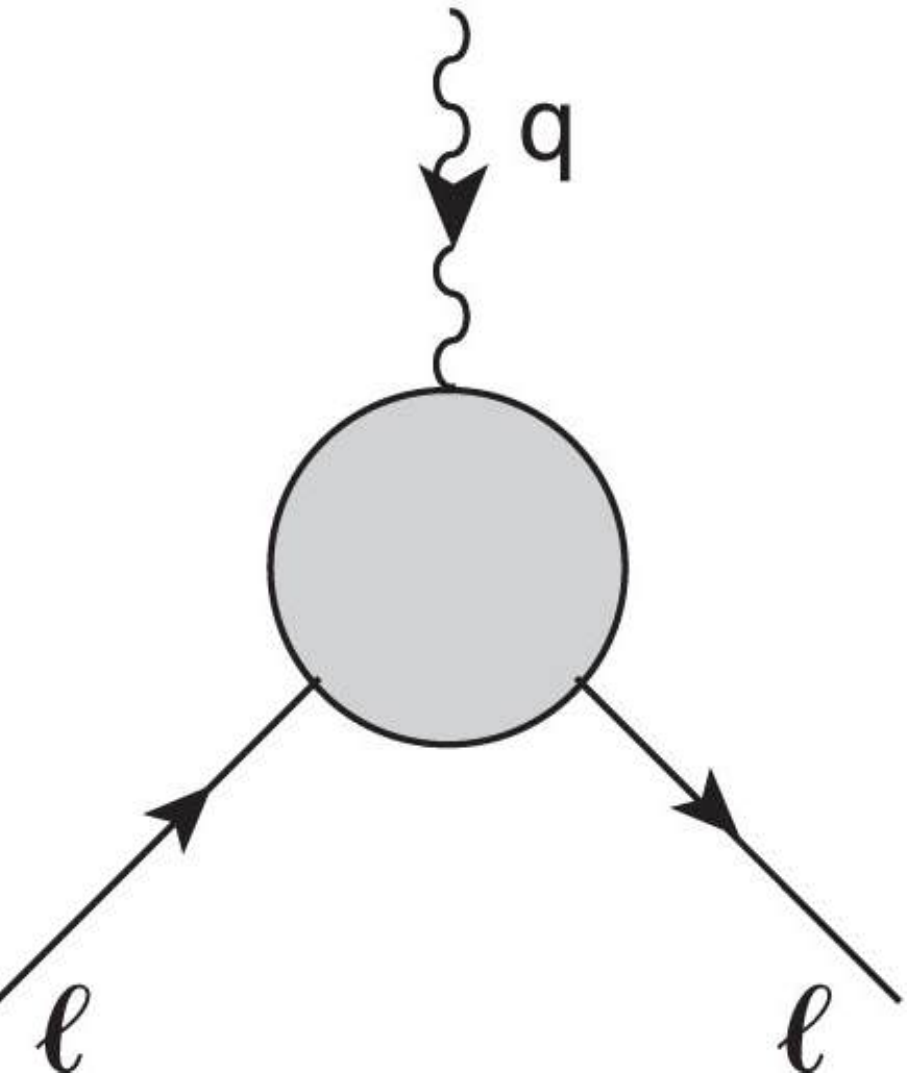}\hspace*{0.5cm} {\purple = }\hspace*{0.5cm}
\includegraphics[width=2.cm]{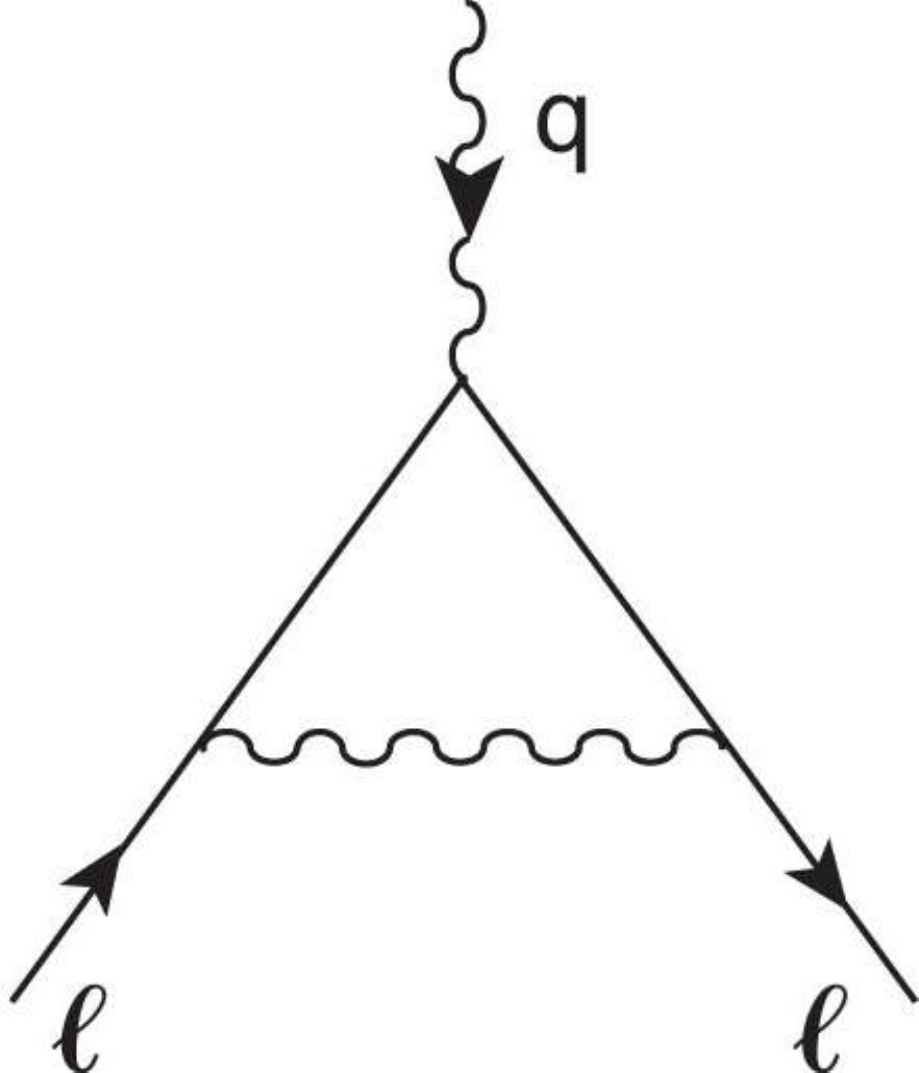}\hspace*{0.5cm}{\purple +
$\cdots$} \\
\hspace*{-1.5cm} a) \hspace*{4cm} b)  \hspace*{3cm} c)

\caption{a) The intrinsic momentum of a self-rotated electron; b) Box diagram representing the quantum corrections to the tree level Dirac equation solution $g=2$; c) First quantum corrections to $g$. }
\end{center}
\label{fig:g-2}
\end{figure}
} 

\d The lepton anomaly is defined as\,:
\beq
a_l\equiv \frac{1}{2} ({ g}-2). 
\eeq
The Dirac equation gives :
$
{ g=2}, 
$
while the first quantum correction in Fig.\,\ref{fig:g-2}c) calculated by Schwinger (1947)
gives\,:
\beq
 a_l= \frac{\alpha}{2\pi}=0.00116...
 \eeq
where $\alpha = 1/137...$ $a_e$ and $a_\mu$ are known to order $10^{-11}$ and are the most precise quantities in Physics. 

\section*{
\b Motivation for analyzing the muon anomaly}

\d  Like in the case of the electron anomaly, one also expects that the analyze of the muon anomaly can provide a stringent test of perturbative QED.
 
\d It is also a good place to study the enhancement of the mass ratio $m_\mu/m_e$ where le leading contributions are log$(m_\mu/m_e)$. 

\d The hadronic contributions are more important here than for the case of the electronand are the main sources of the errors in the determination of $a_\mu$.

\d The wish and the hope is to give constraints  on New Physics beyond the Standard Model via virtual effects from Low-Energy process wich is an alternative or a complement of the BSM physics  search at high-energy accelerators like the LHC@CERN. 

\section*{
\b First Measurements of the muon anomaly}

\d The measurement of the muon anomaly has been initiated in Ref.\,\cite{CASSEL} from a polarized 110 MeV postive muon beam produced from pion decays 
with the result:
\beq
a_\mu^{exp}= (2\pm 7)\times 10^{-4},
\eeq
Later on, Ref.\,\cite{COFFIN} uses a magnetic resonance technique on the  CHBr3 target and found:
\beq
a_\mu^{exp}= (26\pm 9)\times 10^{-4},
\eeq
which agrees within the error  with the QED prediction of $12\times 10^{-4}$.  

\d  Using a precession technique for measuring the ratio of the muon precession frequency to that of the proton in the same magnetic field 
and the lower limit of the muon mass, 
Ref.\,\cite{GARWIN} found the lower limit:
\beq
a_\mu^{exp}\geq (122\pm 8)\times 10^{-5},   
\eeq
Using the same procession technique for for 100 MeV/c muons as a function of storage time t, Ref.\,\cite{CHARPAK} found
\beq
a_\mu^{exp}=(1162\pm5)\times 10^{-6},
\eeq
in agreement of the QED prediction to order $(\alpha/\pi)^2$ : 
\beq
a_\mu^{qed}=1165\times 10^{-6} 
\eeq
\section*{\b Modern measurements of the muon anomaly}
\d The CERN muon storage ring (1961-1976)

A muon storage ring dedicated to a direct measurement of $a_\mu$ via the produced muon precessing spin has been built at CERN with the final average from positive and negative charged muons\,\cite{BAILEY}\,:
\beq
\la a_\mu\ra = (1165924.0\pm 8.5) \times10^{-9},
\eeq
in good agreement with the theoretical estimate within the Standard Model \,\cite{CALMET}\,:
\beq
a_\mu^{th} = (1165920.6 \pm 12.9) \times10^{-9}.
\eeq

\d $a_\mu$ from E821 (BNL) and FNAL

Later on, improved measurements of the muon anomaly have been done by\,:

-- The E821 at Brookhaven National Laboratory (BNL)  from 1997 to 2001 which uses the same techniques as the CERN muon storage ring but with innovative technologies (continuous superconducting magnet, having high
field uniformity, continuous current rather than pulsed,...). The  final report of the E821 (BNL) gives\,\cite{BNL}:
\beq
a_\mu\vert^{exp}_{bnl} = (116592080\pm 63)\times 10^{-11}, 
\label{eq:bnl}
\eeq
which is about $(2.2\sim 2.7)\sigma$ above the SM predictions compiled in\,\cite{BNL}.

-- The most recent measurements of $a_\mu$ from FNAL Runs 1 to 6 \,\cite{FNAL1}\,:
\beq
a_\mu\vert^{exp}_{fnal} = (116592070.5\pm 14.8)\times 10^{-11},
\label{eq:final}
\eeq
\begin{figure}[hbt]
\begin{center}
\includegraphics[width=6cm]{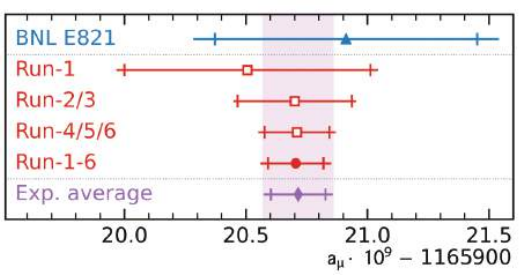}
\vspace*{-0.5cm}
\caption{\scriptsize $a_\mu$ from BNL and FNAL }\label{fig:fnal}
\end{center}
\vspace*{-0.5cm}
\end{figure}  
 improve the BNL result  and  reduce the previous FNAL Run2 error\,\cite{FNAL2} by a factor 1.7, which leads to the BNL and FNAL average\,(see Fig\,\ref{fig:fnal} from \cite{FNAL1})\,:
\beq
\la a_\mu^{exp}\ra = (116592071.5\pm 14.5)\times 10^{-11}.
\eeq
The previous experimental progresses stimulated an improvement of the theoretical predictions since the lowest order Schwinger quantum correction from diagram with an exchanged  virtual photon in Fig.\ref{fig:g-2}c.  

In this reviiew, we first emphasize the estimate of the hadronic contributions to the muon anomaly which is the main source of errors for its Standard Model (SM)  predictions. 

\section{The lowest order hadronic vacuum polarization contribution $a_\mu\vert^{hvp}_{lo}$}
\begin{figure}[hbt]
\begin{center}
\includegraphics[width=4cm]{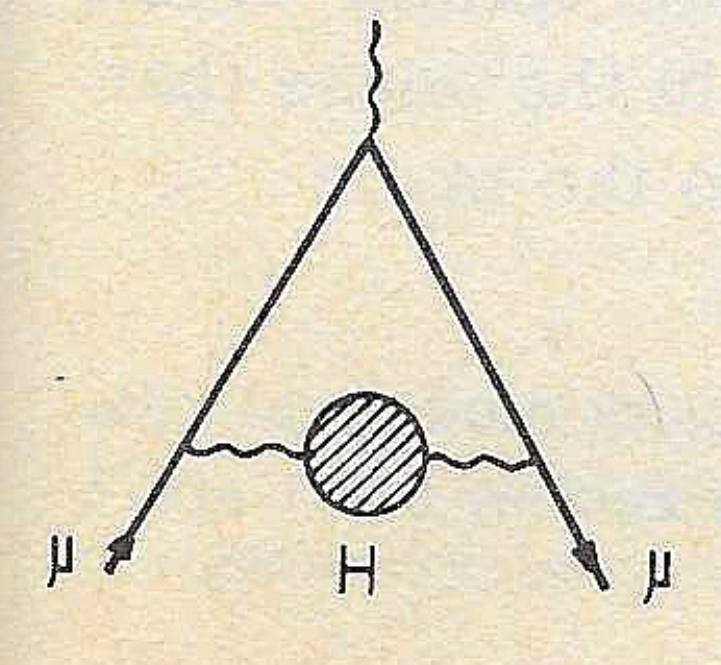}
\caption{\scriptsize Lowest order hadronic vacuum polarization contribution to $a_\mu$ }\label{fig:amu-hvp-lo}
\end{center}
\vspace*{-0.5cm}
\end{figure}

This contribution is given by the diagram in Fig.\,\ref{fig:amu-hvp-lo} where one has  hadrons and quark loops in the photon propagator. 
This contribution can be written in terms of a convolution integral in terms 
Using the analyticity of the spectral function of the electromagnetic 
two-point correlato and its relation to the total cross-section of the $e^+e^-\to$ Hadrons via the optical theorem,
one can express this contribution as a convolution integral \,(see Ref.\,\cite{DERAFAEL})\,:
\beq
a_\mu\vert_{lo}^{hvp}= \frac{1}{4\pi^3}\int_{4m_\pi^2}^\infty dt\, K_\mu(t)\sigma (e^+e^-\to {\rm hadrons})
\label{eq:amu-form}
\eeq
where $K_\mu(t)$ is the QED kernel function\,:
\beq
K_\mu(t)=\int_0^1dx\frac{x^2(1-x)}{x^2+(t/m_\mu^2)(1-x)}.
\label{eq:kmu}
\eeq
$K(t)$  behaves as $1/t$ at low-energy such that the integral is more sensitive to the lowest meson mass contributions thsn to the higher mass ones. 

In the following we estimate $a_\mu\vert^{hvp}_{lo}$ using  the $e^+e^-\to$ Hadrons total cross-section data.  
\section{The $\rho$ meson contribution below 0.93 GeV}
\begin{figure}[hbt]
\begin{center}
\includegraphics[width=11.cm]{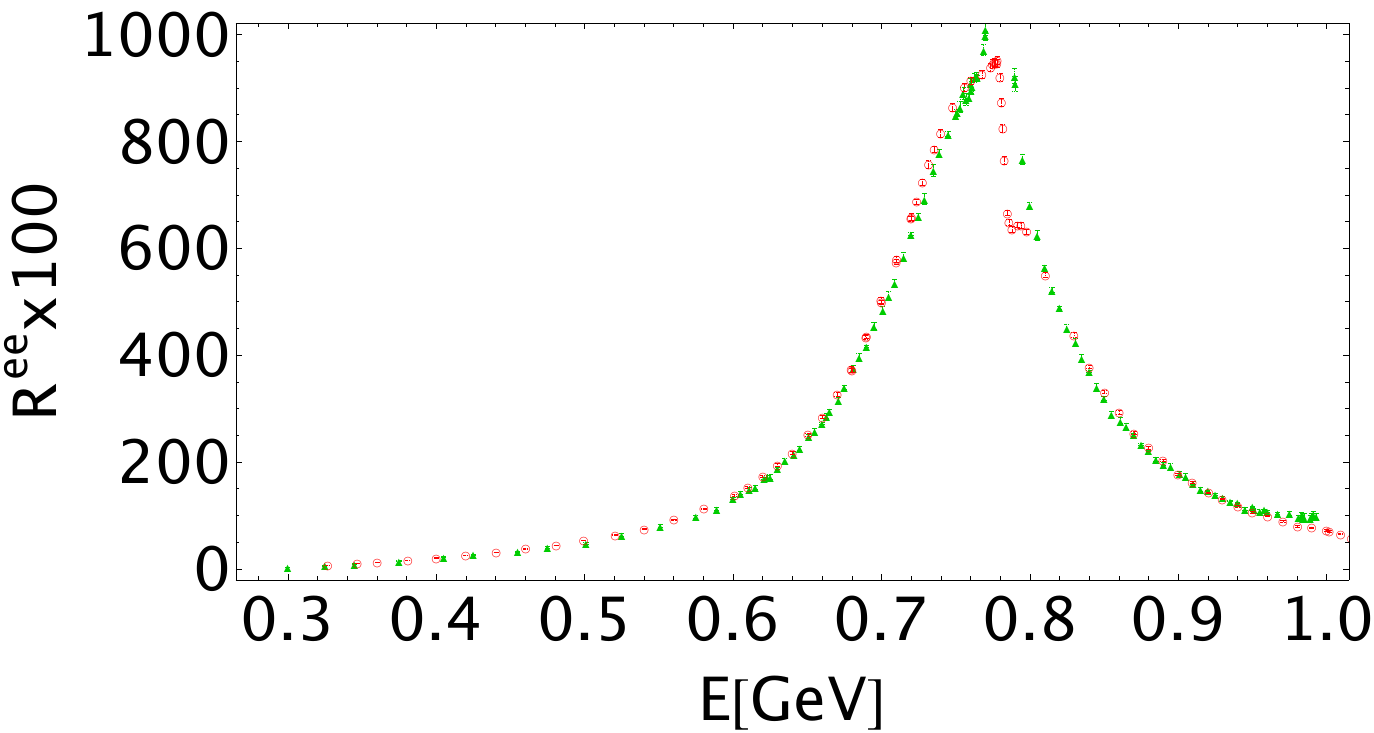}
\caption{\footnotesize  A comparison with the new CMD-3\,\cite{CMD3} data (open red circle) with the PDG22 compilation\,\cite{PDG} (green triangle). } \label{fig:rho}
\end{center}
\vspace*{-0.5cm}
\end{figure}

We are certainly aware about the discrepancy among different previous data for the isovector channel $e^+e^-\to\pi^+\pi^-$ and $\tau$-decay. Then we have estimated this contribution using the most recent CMD-3 data\,\cite{CMD3} which seem to be confirmed by some new preliminary SND data.   We shall work with the ratio of cross-sections\,:   
 \beq
 R^{ee} \equiv \frac{ \sigma (e^+e^-\to {\rm I=1\,hadrons})}{\sigma (e^+e^-\to \mu^+\mu^-)} = \frac{1}{4}\ga 1-\frac{4m_\pi^2}{t}\dr^{3/2} \vert F_\pi\vert^2.
\eeq
where\,: $F_\pi$ is the form factor. The data are shown in Fig.\,\ref{fig:rho} where we compare the new CMD3 data with the one compiled by PDG22. We see a net difference around the $\rho-\omega$ mixing region.  One can also notice that in the regions from 0.4 to 0.6 GeV and 0.68 to 0.77 GeV, the CMD-3 data are systematically above the PDG22 compilation.  
In order to avoid any model-dependent theoretical parametrization of the form factor, we subdivide this region into 5 different sub-regions (see details in Refs.\,\cite{SN1,SN2,SN3,SN4,SN5})\,:
\beq
[2m_\pi\to 0.6],\,\,\,[0.6\to 0.776],\,\,\,[0.776\to 0.786],\,\,\,[0.786 \to 0.810],\,\,\, [0.810\to .993],
\eeq
where the data are fitted with polynomials using the optimized $\chi^2$ Mathematica program {\it Findfit} (see the original fit in Figs. 18 and 19 in Ref.\,\cite{SN5}).  To be conservative, we consider the region spanned by the low and high values of the data.  Our optimal result  is the mean of these two extremal values. We may eventually expect that the results are less accurate than the one from a fit of the correlated data used currently in the literature.

Our fitting procedure are extended to the different  available data below 11.5 GeV where, in some cases, we use a simple Breit-Wigner fit. 

We cut the fit at 0.993 GeV where the ratio R starts to increase indicating the opening of new channels. 
Collecting the results from different regions, we deduce from the CMD-3 data\,:
\beq
a_\mu\vert_{lo}^{hvp}\vert_{cmd-3}(2m_\pi\rar 0.993) = (5109.3\pm 35.2)\times 10^{-11},
\eeq
where 76\% of the contribution comes from the region below the $\rho$-meson mass.

\section{Light $I=1$ isovector mesons  from 0.993 to  1.875 GeV}

\b We fit the data between 0.993 to 1.5 GeV using a simple cubic polynomial interpolation where we substract the $3\pi$ backgrounds
 by using  the suppression 1/9 factor of the isoscalar versus the isovector states from $SU(2)$ symmetry.
 The $\bar KK$ contributions from isoscalar sources are doubly suppressed from  the $SU(3)$ suppression factor and by phase space. The fit is shown in Fig. 9 of \cite{SN1}.    
We obtain:
\beq
a_\mu\vert_{lo}^{hvp}(0.993\rar 1.5) = (354.4\pm 6.7)\times 10^{-11},
\eeq


\b The $\rho'(1600)$ is fitted by a simple Breit-Wigner fit from 1.5 to   1.875 GeV (see Fig. 9b of \cite{SN1}) where we obtain :
\beq
M_{\rho'}=1.6\,{\rm GeV}\,\,\, {\rm and}\,\,\,\Gamma(\rho'\to e^+e^-) = [10.5\,({\rm resp.}\, 9.73)]~{\rm keV}~~~~~~~~\Gamma_{\rho'}^{\rm tot}=  [720\,({\rm resp.}\,694)]\,\rm{MeV},
\eeq
from the high (resp. low) data points. 
We deduce:
\beq
a_\mu\vert_{lo}^{hvp}[\rho'(1600)] = (237.6\pm 5.7)\times 10^{-11},
\eeq

\section{Sum of Light $I=1$ isovector mesons  contributions from $2m_\pi$ to  1.875 GeV}
From the previous analysis, we deduce the total sum of the $I=1$ isovector meson contributions
from $2m_\pi$ to  1.875 GeV:
\beq
a_\mu\vert_{lo}^{hvp}(2m_\pi\rar 1.875) = (5701.3\pm 36.3)\times 10^{-11},
\label{eqq:i=1}
\eeq
which can be compared with previous  different results compiled in Table\,\ref{tab:amu1}. 
\section{The $\omega(780)$  $I=0$ isoscalar meson}
\subsection*{\hspace*{0.5cm} \d The $\omega(780)$ meson contribution below 0.96 GeV}

Instead of the Narrow Width Approximation (NWA) used in Ref.\,\cite{SN5}, we use the $e^+e^-\to3\pi$ data from BABAR\,\cite{BABAR} (most accurate) and BELLE II\,\cite{BELLE} (most recent). We fit the data using some polynomials. The analysis is shown in Figs.10 and 11 of Ref.\,\cite{SN1}.  The fit until 0.99 GeV from which we shall use the full data of $R{e^+e^-}$ compiled by PDG\,\cite{PDG}. The change from 0.96 to 0.99 GeV increases the contribution to $a_\mu$ by $4\times 10^{-11}$. We obtain:
\bea
a_\mu\vert_{lo}^{hvp}(3m_\pi\rar 0.99) &=& (333.7\pm 9.8)\times 10^{-11}
~~~~~{\rm BELLE II} , \nnb\\
&=& (313.3\pm 3.8)\times 10^{-11}~~~~~{\rm BABAR} 
\eea
where the BABAR data leads to a result 3 times more accurate  than the one of  BELLE II.  
In this channel,  about 43\% of the contribution comes from the region below the $\omega$-meson mass.
This result can be compared with the one from NWA\,\cite{SN5}\,:
If instead, we use a NWA as in Ref.\,\cite{SN5}, one uses\,:
\beq
a_\mu^{\omega}\vert^{hvp}_{l.o}= \frac{3}{\pi}\frac{\Gamma^{ee}_\omega}{M_\omega} K_\mu(M_\omega^2)\simeq (417\pm 14)\times 10^{-11},
\eeq
where the relative large error is due to the inaccurate value of the $\omega\to e^+e^-$ decay width.  The larger value of $a_\mu$ from the NWA than the one  from the Breit-Wigner cutted at 0.96 GeV may be due to the fact  that the NWA does not take into account the effect of the tail of the $\omega$ width.  

We make a tentative average of the three results by multiplying the BABAR error by a conservative factor 2. In this way, we obtain the mean:
\beq
a_\mu\vert_{lo}^{hvp}(3m_\pi\rar 0.99) = (335.9\pm 5.5)\times 10^{-11},
\label{eq:amu-3pi}
\eeq
which we consider as a final estimate quoted in Table\,\ref{tab:amu1}. 


The values of $a_\mu$ quoted by BABAR $(459\pm 6)\times 10^{-11}$  and BELLE II $(481\pm 11)\times 10^{-11}$ are expected to be larger as they include all $3\pi$ contributions below 1.8 GeV.
\section{The $\phi(1020)$ $I=0$ isoscalar meson}
We estimate its contribution using a NWA. In this way, we obtain\,\cite{SN5}:
\beq
a_\mu\vert_{lo}^{hvp} \vert^{\phi}_{nwa}= (389.6\pm 4.6)\times 10^{-11}.
\eeq
Once we have isolated the $\phi$-meson contributions, we estimate the contributions from 0.96 to 1.875 GeV using the data compiled by PDG on the ratio $R^{ee}$.  The analysis has been done in ref.\,\cite{SN5}. 

\section{Light $I=0$ Isoscalar mesons from 0.96 to 1.875 GeV}
\subsection*{\hspace*{0.5cm} \b Data from 0.99 to 1.5 GeV}
We consider the $I=0$ contribution  substracted  in the $I=1$ analysis of the previous sections by using  the $SU(3)$ relation. We fit the data using a polynomial. We obtain\,\cite{SN5}\,: 
\beq
a_\mu\vert_{lo}^{hvp}(0.99\rar 1.5) = (44.3\pm 0.8)\times 10^{-11},
\eeq
\subsection*{\hspace*{0.5cm} \b Data from 1.5 to 1.875 GeV}
We consider the contributions of the $\omega(1650)$ and $\phi(1680)$. We use the parameters deduced from PDG22\,\cite{PDG}:
\bea
\omega(1650)&:&  M_\omega=(1670\pm 30)\,{\rm MeV},\,\,\,\,\,\,        \,\,\,\,\,  \,\,\,\Gamma^{ee}_\omega=1.35\pm 0.14\,{\rm keV},\,\,\,\,\,\,  \,\,\,\,\,\,  \,\,\Gamma^{h}_\omega=315\pm 35\,{\rm MeV} \nnb\\
\phi(1680)&:&  M_\phi=(1680\pm 20)\,{\rm MeV},\,\,\,\,\,\,   \,\,\,     \,\,\,\,\,  \,\,\,\Gamma^{ee}_\phi=0.18\pm 0.06\,{\rm keV},\,\,\,\,\,\,  \,\,\,\,\,\,  \,\,\Gamma^{h}_\psi=150\pm 50\,{\rm MeV}.
\eea

We use a Breit-Wigner for describing the $\omega(1650)$ and $\phi(1680)$ meson. We obtain\,:
\bea
a_\mu\vert_{lo}^{hvp}[\omega (1650)] &=& (24.3\pm 0.1)\times 10^{-11},\nnb\\
a_\mu\vert_{lo}^{hvp}[\phi (1680)] &=& (1.8\pm 0.9)\times 10^{-11},
\eea
\section{Sum of Light $I=0$ isoscalar mesons  contributions from $3m_\pi$ to  1.875 GeV}
Adding the sum of previous $I=0$ isoscalar mesons contributions, we deduce:
\beq
a_\mu\vert_{lo}^{hvp}(3m_\pi\rar 1.875) = (795.9\pm 7.7)\times 10^{-11},
\eeq
compiled in Table\,\ref{tab:amu1}. 

\section{$I=1\oplus 0$ light mesons from 1.875 to 3.68 GeV}
\subsection*{\hspace*{0.5cm} \b Data from $1.875\rar 2$ GeV}
To get this contribution, we divide the data into three regions:
\beq
[1.875\to 1.91],\,\, [1.91,1.96],\,\,[1.96,2.00] 
\eeq
whih we fit with polynomials (see Fig. 20 of Ref.\,\cite{SN5}). 
We obtain (see details in Table\,\ref{tab:amu1})\,:
\beq
a_\mu\vert_{lo}^{hvp}(1.875\rar 2) = (46.6\pm 1.0)\times 10^{-11}.
\eeq
\section { QCD continuum from $2\rar 3.68$ GeV}
\begin{figure}[hbt]
\begin{center}

\includegraphics[width=12cm]{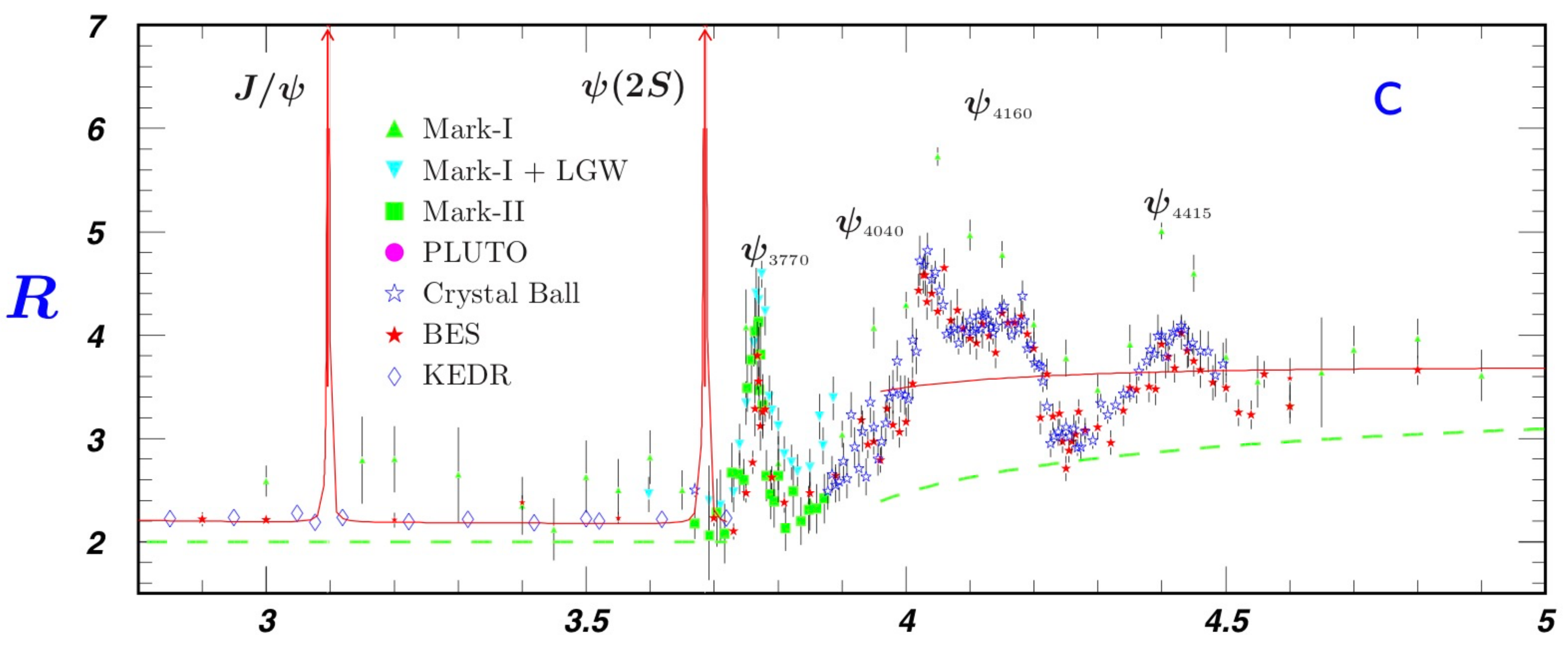}
\caption{\footnotesize  $e^+e^-\to$ Hadrons PDG data in the charmonium region from 2 to 5 GeV from PDG\,\cite{PDG}.  The broken green line is the naive parton model prediction. The continuous red line is the QCD prediction including 3-loop PT corrections.} \label{fig:charm}
\end{center}
\vspace*{-0.5cm}
\end{figure}

The data in this region are well fitted by the QCD expression of the spectral function for 3 flavours as one can see in Fig.\,\ref{fig:charm} from PDG\,\cite{PDG}. 

We use the PT  QCD expression  given in Section 2 of Ref.\,\cite{SN5} to which we add the Non-Perturbative contributions quoted in this Section 2 and the HO $m_q^2$ and $m_q^4$ quark mass corrections given in Eq.\,72  of Ref.\,\cite{SN5}. These QCD expressions will also be used for the charmonium and bottomium channels discussed in the next section. We obtain:
\beq
a_\mu\vert_{lo}^{hvp}(2\rar 3.68) = (247.2\pm 0.3)\times 10^{-11},
\eeq

\section{Final sum of the Light $I=1\oplus 0$ light mesons from $2m_\pi$ to 3.68 GeV}
Adding the different contributions of the $I=0$ and 1 light mesons below 2 GeV $\oplus$ the 
corresponding QCD continuum from 2 to 3.68 GeV, we obtain:
\beq
a_\mu\vert_{lo}^{hvp}(2m_\pi\rar 3.68) = (6791.0\pm 37.1)\times 10^{-11}.
\eeq

\section{Charmonium from $J/\psi$ to 10.5 GeV}

\subsection*{
\b Narrow resonances}
We estimate the contributions of $\psi(1S),  \psi(2S)$ and $\psi(3773)$ using a NWA. 
 The result is given in Table\,\ref{tab:amu1}.
\begin{figure}[H]
\begin{center}
\includegraphics[width=12cm]{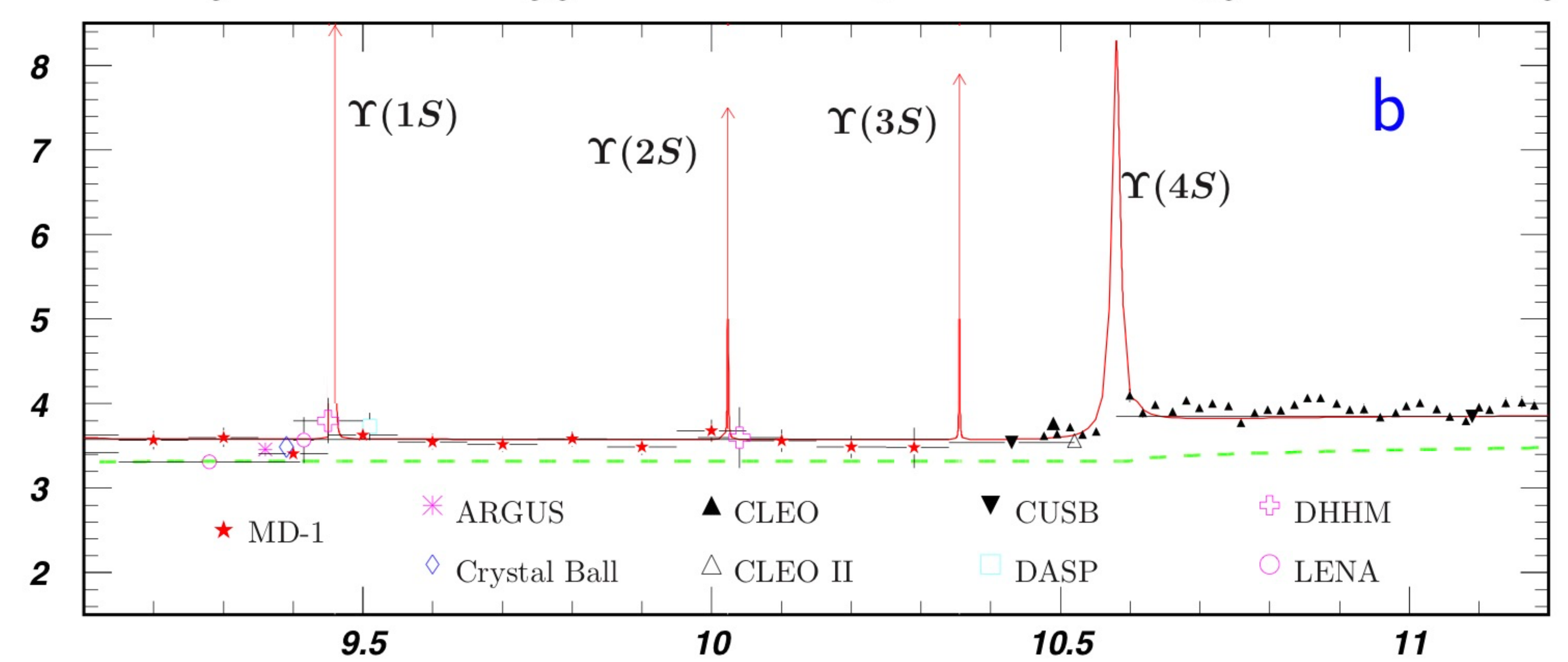}
\caption{\footnotesize  $e^+e^-\to$ Hadrons data in the bottomium region from 9 to 11.5 GeV from PDG\,\cite{PDG}. The broken green line is the naive parton model prediction. The continuous red line is the QCD prediction including 3-loop PT corrections. } \label{fig:bottom}
\end{center}
\vspace*{-0.5cm}
\end{figure} 

\subsection*{
\b Region from 3.68 to 4.55 GeV}
We divide this region into 5 subregions :
\beq
[3.68\to 3.86],\, [3.86 \to 4.09],\, [4.09\to 4.18],\,  [4.18\to 4.30],\, [4.30\to 4.55],
\eeq
 as shown in Fig.\,22 of Ref.\,\cite{SN5}. The region from 3.68 to 3.86 GeV is better fitted using a Breit-Wigner while, for the others,  we use polynomials. The different contributions are shown in Table\,\ref{tab:amu1}. 
\subsection*{\hspace*{0.5cm}\d Region from 4.55 to 10.50 GeV }
As shown in Figs.\,\ref{fig:charm} and \ref{fig:bottom}, the data without resonance peaks are well fitted by QCD for 4 flavours\,\cite{PDG}. We add to the previous QCD expressions the charm contributions including $m_c^2$ and $m_c^4$  mass corrections. We use\,: $\overline{m}_c(\overline{m}_c)=$ 1266(6) MeV\,\cite{SNREV1}. The contribution is given in Table\,\ref{tab:amu1}. 
\section{$a_\mu\vert^{hvp}_{l.o}$ from bottomium}
\b We use a NWA to estimate the contributions of $\Upsilon(1S), \Upsilon(2S), \Upsilon(3S)$ and $\Upsilon(4S)$. 
The results are reported in Table\,\ref{tab:amu1}. 
\subsection*{\b QCD continuum contribution from  10.59 GeV to $2m_t$}
We add the $b$-quark contribution to the previous QCD expression where $b$-quark mass corrections to order $a_s^3\overline{m}_b^2/t$ and $a_s^2\overline{m}_b^4/t^2$ are included. We use the running mass\,\: $\overline{m}_b(\overline{m}_b)=$ 4202(8) MeV\,\cite{SNREV1}.
For the analysis, we consider the region from 10.59 MeV to $2m_t$ just after the $\Upsilon(4S)$ where the QCD continuum is expected to smear the  $\Upsilon(10860, 11020)$  and some eventual  higher resonances. The result is given in Table\,\ref{tab:amu1}. 
\section{QCD continuum contribution from  $2m_t\to\infty$}
Due to the heaviness of the top quark mass, we expect that the approximate Schwinger formula near the $\bar tt$ threshold
gives  a much better description of the spectral function up to order $\alpha_s$\,:
\beq
R_t^{ee}=\frac{4}{3} v\frac{(3-v^2)}{2}\Big{[} 1+\frac{4}{3}\alpha_sf(v)\Big{]},
\eeq
with:
\beq
f(v)=\frac{\pi}{2\, v}-\frac{(3+v)}{4}\ga \frac{\pi}{2}-\frac{3}{4\pi}\dr\,:\,\,\,\,\,\, v=\ga 1-\frac{m_t^2}{t}\dr^{1/2}.
\eeq
Here $m_t$ is the on-shell top quark mass which we fix to be\,\cite{PDG}\,\footnote{We should note that the definition of the top quark mass from different experiments is still ambiguous.}: 
\beq
m_t\approx \overline{m}_t(\overline{m}_t)=(172.7\pm 0.3)\,{\rm  GeV},
\label{eq:mt}
\eeq
 from some  direct measurements.
We add to this expression the  one due to $\alpha_s^2$ and $\alpha_s^3$, given in Ref.\,\cite{SN5}, within the $\overline {MS}$-scheme.  For this $\overline {MS}$-scheme expression  in terms of running mass, we need the value of the RGI top mass:
\beq
\hat m_t=(254\pm 0.4)\,{\rm GeV},
\eeq
deduced from Eq.\,\ref{eq:mt} (for reviews, see e.g.\,\cite{SNB1,SNB2}).
Adding the above expressions to the ones in the previous sections, we obtain the result given in Table\,\ref{tab:amu1}. 

   {\footnotesize
\begin{table}[H]
\vspace*{-1.6cm} 
\setlength{\tabcolsep}{1.pc}
  \begin{center}
    { \footnotesize
  \begin{tabular}{| llll |}
\hline
\rowcolor{yellow}\boldmath$\sqrt{t}$\,\bf [GeV] &\boldmath$a_\mu\vert^{hvp}_{l.o}\times 10^{11} $&\boldmath$a_\tau\vert^{hvp}_{l.o}\times 10^{9} $&$\Delta\alpha^{(5)}_{had}(M_Z^2)\times 10^5$\\ 
 \hline 

{\bf Light I=1}&&&\\
$\rho(2m_\pi\to 0.60)$&$1069.0\pm 22.1$&$85.1\pm 0.6$&$12.38\pm 0.08$\\
$\rho(0.60\to 0.778)$&$2795.9\pm 26.0$&$943.7\pm10.3$&$182.87\pm 1.97$ \\
$\rho(0.778\to 0.993)$&$1244.4\pm 8.6$&$1797.3\pm 20.3$&$117.82\pm 0.98$\\
$0.993\to 1.5$&$354.4\pm 6.7$&$228.2\pm 4.1$&$67.3\pm 1.14$\\
$1.5\to 1.875$&$237.6\pm 5.7$&$206.8\pm 4.9$&$80.11\pm 1.9$\\

{\it Total Light I=1 ($\leq 1.875$)}&$\it 5701.3\pm 36.3$&$\it 2453.3\pm 11.9$&$\it 483.3\pm 4.4$\\
{\bf Light I=0}&&&\\
$3\pi: ~3m_\pi\to 0.99$&$335.9\pm 5.5$&$163.3\pm 5.3$&$33.6\pm 1.1$\\
$\phi$ (NWA) &$389.6\pm 4.6$&$20.6\pm 0.2$&$51.2\pm 0.6$\\
$0.993\to 1.5$ & $44.3\pm 0.8$&$28.5\pm 0.5$&$8.4\pm 0.1$\\
$\omega(1650)$(BW)&$24.3\pm 0.1$&$16.7\pm 0.1$&$5.2\pm 0.1$ \\
$\phi(1680)$(BW)&$1.8\pm 0.9$&$1.3\pm 0.6$&$0.4\pm 0.2$ \\
{\it Total Light I=0 ($\leq 1.875$)}&$\it 795.9\pm 7.3$&$\it 230.4\pm 5.3$&$\it 99.4\pm 1.3$\\
{\bf Light I=\boldmath $0\, \oplus \,1$}&&&\\
$1.875\to 1.913$ & $14.7\pm 0.7$&$14.4\pm 0.7$&$6.3\pm 0.3$\\
$1.913\to 1.96$ & $17.6\pm 0.6$&$17.5\pm 0.6$&$7.9\pm 0.3$\\
$1.96\to 2$ & $14.3\pm 0.5$&$14.5\pm 0.5$&$6.7\pm 0.2$\\
$2\to 3.68$:  QCD $(u,d,s)$ & $247.2\pm 0.3$&$308.3\pm 0.5$&$202.8\pm 0.5$\\
{\it Total Light I=$0\, \oplus \,1$ ($1.875\to 3.68$)}&$\it 293.8\pm 1.1$&$\it 354.7\pm 1.2$&$\it 223.7\pm 0.7$\\
\rowcolor{greenli}{ Total Light I=$0\, \oplus \,1$ ($2m_\pi\to3.68$)}&$ 6791.0\pm 37.1$&$3038.4\pm 24.5$&$806.4\pm 4.6$\\
{\bf Charmonium} &&&\\
$J/\psi (1S)$ (NWA) &$65.1\pm 1.2$&$92.7\pm 1.8$&$73.5\pm 1.4$\\
$\psi (2S)$ (NWA) &$16.4\pm 0.6$&$26.0\pm 0.9$&$26.1\pm 0.8$\\
$\psi (3773)$ (NWA) &$1.7\pm 0.1$&$2.7\pm 0.2$&$2.9\pm 0.2$\\
{\it Total $J/\psi (NWA)$} &$\it 83.2\pm 1.4$&$\it 121.4\pm 2.0$&$\it 102.5\pm 1.6$\\
$3.69\to 3.86$ & $11.4\pm 1.0$&$18.3\pm 1.6$&$19.0\pm 1.6$\\
$3.86\to4.094$ & $16.6\pm 0.5$&$27.5\pm 0.8$&$30.9\pm 0.9$\\
$4.094\to4.18$ & $6.6\pm 0.2$&$11.2\pm 0.3$&$13.2\pm 0.4$\\
$4.18\to4.292$ & $6.5\pm 0.2$&$11.2\pm 0.4$&$13.7\pm 0.5$\\
$4.292\to4.54$ & $11.8\pm 0.6$&$20.7\pm 0.8$&$26.8\pm 1.1$\\
$4.54\to10.50$:  QCD $(u,d,s,c)$ & $92.0\pm 0.0$&$186.2\pm 0.0$&$458.7.3\pm 0.1$\\
{\it Total Charmonium ($3.69\to 10.50$)}&$\it 145.1\pm 1.3$&$\it 275.6\pm 2.0$&$\it 564.9\pm 2.2$\\
\rowcolor{greenli}{ Total Charmonium }&$ 228.1\pm 1.9$&$396.5\pm 2.8$&$664.8\pm 2.7$\\
 {\bf Bottomium} &&&\\
$\Upsilon (1S)$ (NWA) &$0.54\pm 0.02$&$1.25\pm 0.07$&$5.65\pm 0.29$\\
$\Upsilon(2S)$ (NWA) &$0.22\pm 0.02$&$0.51\pm 0.06$&$2.54\pm 0.29$\\
$\Upsilon (3S)$ (NWA) &$0.14\pm 0.02$&$0.33\pm 0.04$&$1.77\pm 0.23$\\
$\Upsilon(4S)$ (NWA) &$0.10\pm 0.01$&$0.23\pm 0.03$&$1.26\pm 0.16$\\
$\Upsilon(10.86\,\oplus\,11)$ (NWA) &$0.1\pm 0.06$&$0.20\pm 0.06$&$1.67\pm 0.39$\\
{\it Total Bottomium (NWA)}&$\it1.0\pm 0.1$&$\it 2.3\pm 0.1$&$\it 11.2\pm 0.5$ \\
$Z-pole $&-&-&29.2\,\cite{YND17}\\
$10.59\to2m_t$\,:  QCD $(u,d,s,c,b)$ & $22.4\pm 0.3$&$57.5\pm 0.1$&$1282.9\pm 1.2$\\
\rowcolor{greenli}{Total Bottomium }&$ 23.4\pm 0.3$&$59.8\pm 0.1$&$1323.3\pm1.3$\\
$2m_t\to\infty$\,: QCD $(u,d,s,c,b,t)$&0.03&0.08&-28.2\\
\hline
\rowcolor{yellow}\bf Total sum &\boldmath$7042.5\pm 37.2$&\boldmath$3494.8\pm 24.7$&\boldmath$2766.3\pm 4.5$\\

   \hline
 
\end{tabular}}
 \caption{ 
  $a_\mu\vert^{hvp}_{l.o}$, $a_\tau\vert^{hvp}_{l.o}$  and $\Delta \alpha(M^2_z)$ within our parametrization of the compiled PDG\,\cite{PDG} $\oplus$ the recent CMD-3\,\cite{CMD3} on $2\pi$ $\oplus$ Babar\,\cite{BABAR} and new Belle2\,\cite{BELLE} $3\pi$ data. The contributions of $\pi^0\gamma$ and $\eta\gamma$ are not included in this Table. We have not updated the estimate of $a_\tau\vert^{hvp}_{l.o}$ and $\Delta \alpha(M^2_z)$ as the effect of the new fit is almost negligible.}\label{tab:amu1} 
 \end{center}
\end{table}
} 

\section{Hadronic Vacuum Polarization (HVP) Contributions from $e^+e^-\to$ Hadrons}
 \subsection*{\b LO Vacuum Polarization\,: $a_\mu\vert^{hvp}_{lo}\times 10^{11}$}
 Data of CMD3 on $2\pi$ $\oplus$ the inclusive $e^+e^-$ ratio of cross-section $\oplus$ HO PT QCD and HO quark mass corrections have been used in SN2023. New Belle 2 data on $3\pi$ has been used in SN2025 in addition to the data in SN2023. One can notice that the central value is stable with time while the error has decreased by a factor 22 afer 50 years thanks to the amount of experimental efforts done along  this period. 
\vspace*{-0.5cm}
\begin{table}[hbt]
 \setlength{\tabcolsep}{1pc}
\hspace*{0.5cm}
\begin{tabular}{ccccc}
&\\
\hline
SN1976-78\cite{CALMET,SN76,SN78}&\blueda SN2001\,\cite{SN01}& \oliva SN2023 &\purple SN2025   \\
\hline
$7020\pm 800$&\blueda$7021\pm 76$&{\oliva $7036.5\pm 36.9$}&\purple $7042.5\pm 37.2$\\  
\hline
\end{tabular}

\end{table}
\begin{figure}[hbt]
\vspace*{-2.6cm}
\hspace*{13.5cm}\includegraphics[width=3cm]{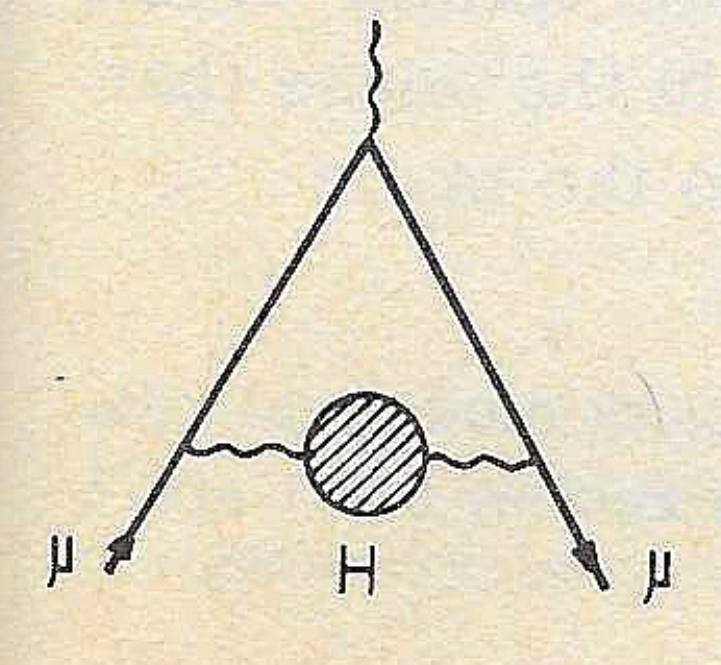}
\end{figure} 
\vspace*{-1.2cm}
\subsection*{\b Contributions of $e^+e^-\to P \gamma$\,\,($P\equiv  \pi^0,\eta$) to $a_\mu^{P\gamma}\vert^{hvp}_{lo}\times 10^{11}$ }
\vspace*{-0.25cm}
 \setlength{\tabcolsep}{2.3pc}
 
   {
   \normalsize
\begin{tabular}{lll}
&\\
\hline
\oliva KNTZ19\,\cite{NOMURA}& DHZ20\,\cite{DAVIER2} &\purple  Mean\\
\hline
\oliva $51.6\pm 1.03$& $50.6\pm 1.02$&\purple $51.05\pm 0.72$\\
\hline
\end{tabular} 
}
\vspace*{-2.2cm}
\begin{figure}[hbt]
\begin{flushright}
\hspace*{12.5cm}\includegraphics[width=3cm]{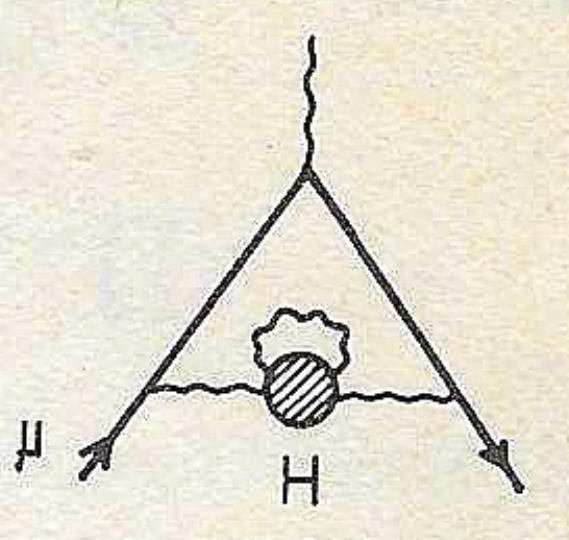}
\end{flushright}
\end{figure}  
\vspace*{-1.5cm}

\subsection*{\b Comparison of different   estimates of $a_\mu\vert^{hvp}_{lo}\times 10^{11}$ including $e^+e^-\to \pi^0,\eta \gamma$}
As mentioned by Ref.\,\cite{IGNATOV}, the CMD3 data do not include the $e^+e^-\to \pi^0\gamma+\eta\gamma$
 process.  Then, we add to the previous determinations this contribution which become\,:
\vspace*{-0.5cm}
\begin{table}[H]
 \begin{center}
 \setlength{\tabcolsep}{0.35pc}
\begin{tabular}{lllllll}
&\\
\hline
\purple This work 2025&\oliva KNT19\,\cite{NOMURA}& DHMZ20\,\cite{DAVIER2}&J26\,\cite{JEGER26}&D9\,$\oplus$\,M24\cite{MASJUAN}&
\blueda [WP25]\,\cite{WP25}&\blueda [BMW20]\,\cite{BMW}\\
\hline
\purple$7093.6\pm 37.2$&\oliva  $6927.8\pm24.2$& $6940\pm 40$&$7023\pm 42$  &$7047\pm 43$& \blueda$7132\pm61$&  \blueda $7075\pm 55$ \\
\purple CMD3 $\oplus \,e^+e^-$ incl.
&\oliva  $e^+e^-$ excl.&$e^+e^-$ excl.&$\rho^0$-$\gamma$ mix.&$\tau$-decays&  \blueda Mean Latt.& \blueda Precise Latt.\\
\hline
\end{tabular} 
\caption{Comparison of some estimates of $a_\mu\vert^{hvp}_{lo}$.}
\end{center}
\label{tab:amu-hvp-lo}
\end{table}
\vspace*{-0.8cm}
\begin{figure}[H]
\begin{center}
\hspace*{0.cm}\includegraphics[width=8cm]{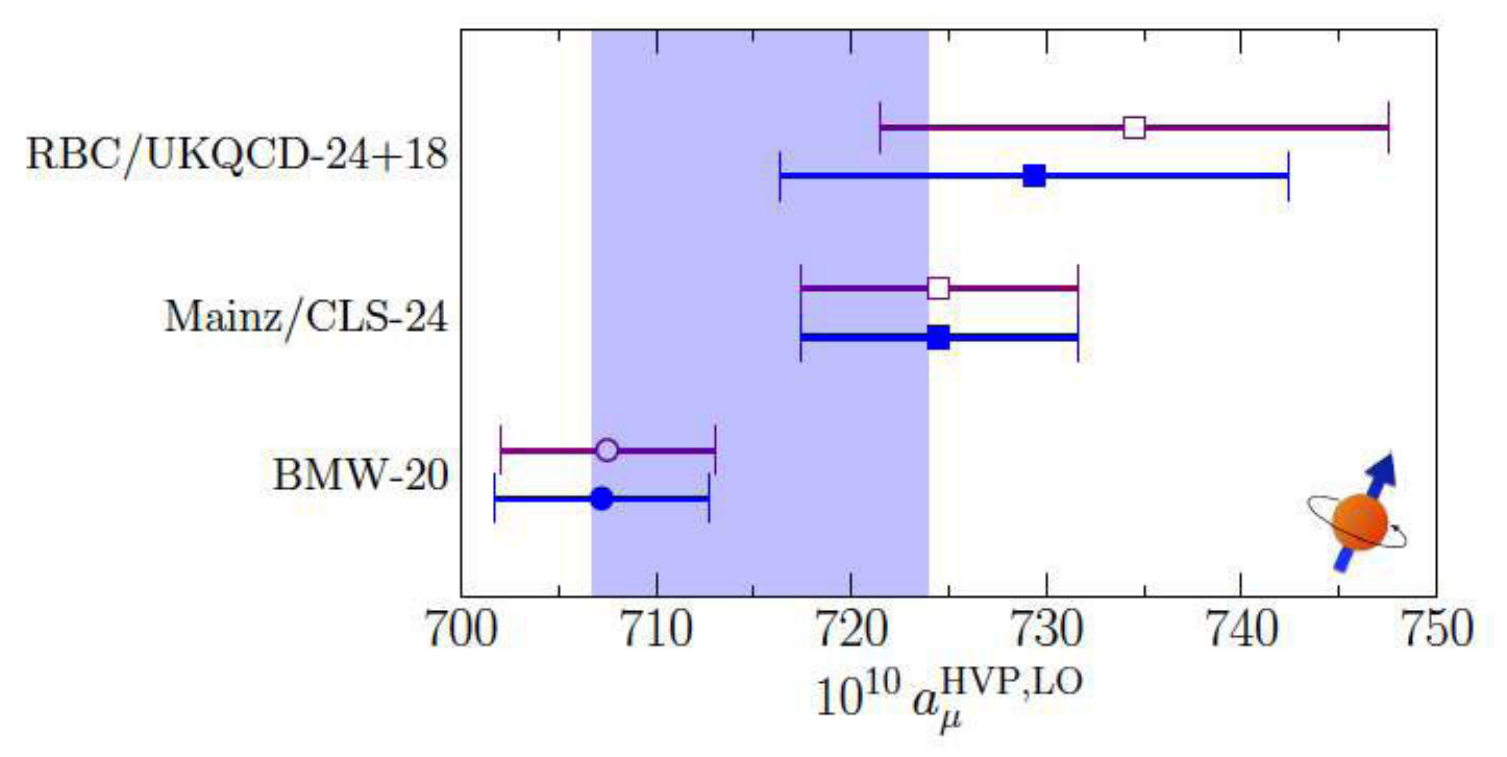}
\vspace*{-0.5cm}
\caption{Some lattice calculations of $a_\mu\vert^{hvp}_{lo}$ compiled by Ref.\,\cite{WP25}.}
\end{center}
\label{fig:latt-hvp}
\end{figure}  
\vspace*{-1.cm}
\subsection*{\b Some comments on the   estimates of $a_\mu\vert^{hvp}_{lo}$}

The results of Refs.\,\cite{NOMURA}  and \cite{DAVIER2} using exlusive $e^+e^-\to$ Hadrons and $\tau$-deay data are  
are respectively lower by\,:
\beq
 \Delta a_\mu\vert^{hvp}_{lo}=(166\pm 31)\times 10^{-11}\,\,{\rm and}\,\,(154\pm 39)\times 10^{-11}
 \label{eq:comparison} 
 \eeq
  than our results. A sharp comparison with these works cannot be easily done  as each authors choose different cuts when connecting different data.  

\subsection*{  \hspace*{0.5cm} 
\d Comparison with Ref.\,\cite{NOMURA} from $e^+e^-\to$ Hadrons  data  below 1.937 GeV}
-- To track the origin of the discrepancy, we compare the total sum of the contributions of the exclusive modes of the light meson channels up to 1.937 GeV from Ref.\,\cite{NOMURA} which are:
\beq
a_\mu\vert^{hvp}_{lo}(\leq 1.937) =(6393\pm 23)\times 10^{-11}
\eeq
to the sum of our estimate until 1.875 GeV given in Table\,\ref{tab:amu1} to which we add the   $e^+e^-\to P\gamma$ contribution:
\beq
a_\mu\vert^{hvp}_{lo}(\leq 1.875) =(6497\pm 37\oplus 51\pm 0.7)\times 10^{-11}.
\label{eq:amu18}
\eeq
The difference between the two determinations is  $(155\pm 31)\times 10^{-11}$ which is approximately equal to the one from the total sum.  

-- By comparing our result up to 0.993 GeV with the one of \cite{NOMURA}, we notice that  approximately $98\times 10^{-11}$ comes from the $2\pi$ channel where from 0.4 to 0.6 GeV and 0.68 to 0.77 GeV, the CMD-3 data are systematically above the PDG22 compilation (see Fig.\,\ref{fig:rho}).

\subsection*{ \hspace*{0.5cm} 
\d $e^+e^-\to$ Hadrons data below $(1.8\sim 1.875)$  GeV \,\cite{DAVIER2}}
We track the difference between Ref.\,\cite{DAVIER2} and ours by examining two regions\,:

-- The results of Ref.\,\cite{DAVIER2} below $(1.8\sim 1.875)$ GeV:
$
a_\mu\vert^{hvp}_{lo}(\leq 1.8)= (6356\pm 40)\times 10^{-11}
$
is $(192\pm 39)\times 10^{-11}$ lower than our result in Eq.\,\ref{eq:amu18}. 

-- The ones from 1.8 to 3.7 GeV:
$
a_\mu\vert^{hvp}_{lo}(1.8\to 3.7)$\,\cite{DAVIER2}$ = (334.5\pm 7.1)\times 10^{-11},
$
are higher  than ours by $(41\pm 3)\times 10^{-11}$.

-- For the total sum of $a_\mu\vert^{hvp}_{lo}$ in all regions, the two differences tend to compensate and lead to the final underestimate of about $154\times 10^{-11}$ quoted in Eq.\,\ref{eq:comparison}.


\subsection*{\b Hadronic vacuum polarization  to higher orders}
\begin{figure}[hbt]
\begin{center}
\includegraphics[width=4.4cm]{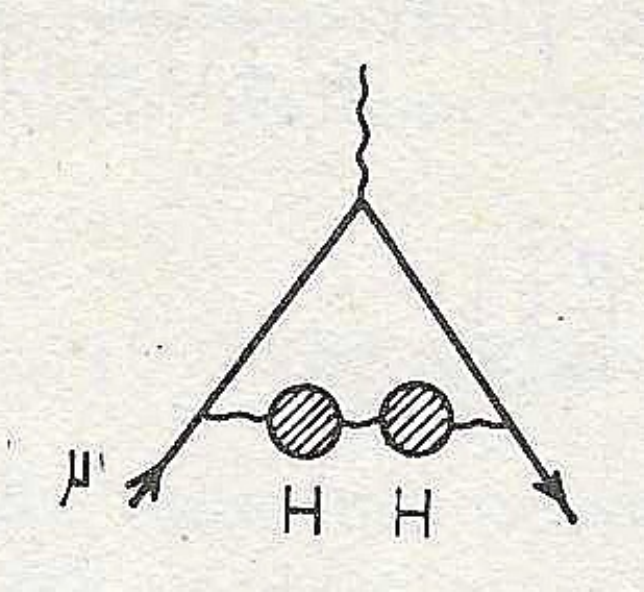}
\includegraphics[width=4.5cm]{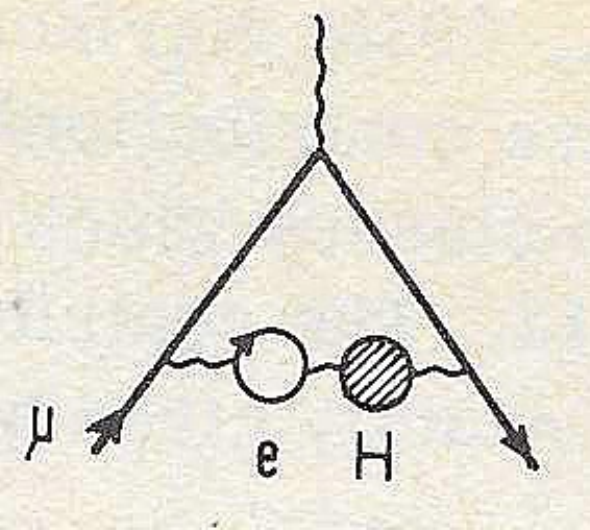}
\includegraphics[width=4.6cm]{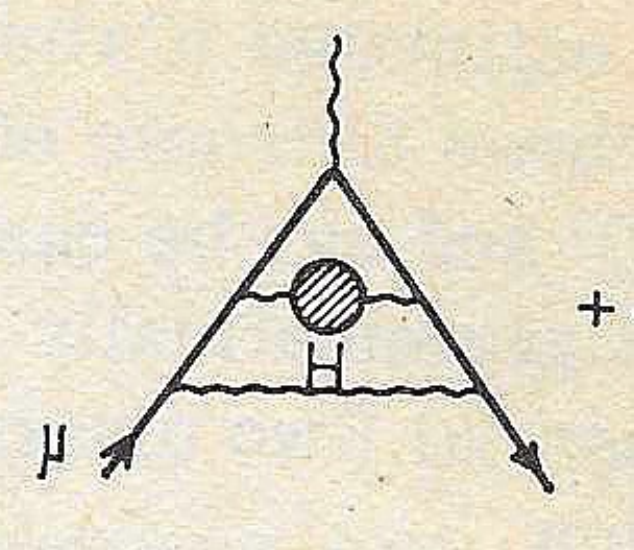} 
\caption{Some typical contributions to $a_\mu\vert^{hvp}_{nlo}$.}
\end{center}
\label{fig:amu-vp-nlo}
\end{figure}  
\vspace*{-1cm}
The NLO and N2LO contributions shown in Fig.\,\ref{fig:amu-vp-nlo} are known and reviewed in Ref.\,\cite{CALMET,WP25}. The estimate from $e^+e^-\to $ Hadrons data is\,\cite{WP25}\,:
\beq
\Delta a_\mu\vert^{hvp}_{nlo}= -(99.6\pm 1.3)\times 10^{-11},~~~~~~~~~~~~~~\Delta a_\mu\vert^{hvp}_{n2lo}= (12.4\pm 1.0)\times 10^{-11},
\label{eq:hvp-ho}
\eeq

\section{The Hadronic Light by light scattering contribution}
\begin{figure}[hbt]
\begin{center}
\includegraphics[width=4.cm]{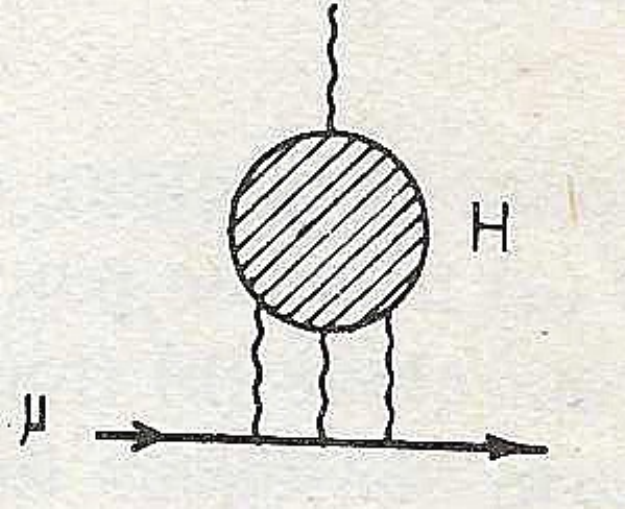}
\caption{LO hadronic light by light (hlbl) scattering contribution to $a_\mu\vert^{hlbl}_{lo}$.}
\label{fig:hlbl}
\end{center}  
\end{figure}

The LO hadronic light by light (hlbl) scattering contribution is represented by the Feynman diagram in Fig.\,\ref{fig:hlbl}. Some estimates of different meson contributions are reviewed in Ref.\,\cite{WP25} where the $I=0$ scalar meson contribution has been estimated in Ref.\,\cite{KNECHT}. They read:
\beq
a_\mu\vert^{hlbl}_{lo}= (103.3\pm 8.8)\times 10^{-11},~~~~~~~~~~~~~~\Delta a_\mu\vert^{hlbl}_{nlo}= (2.6\pm 0.6)\times 10^{-11},
\label{eq:hlbl-pheno}
\eeq
One can notice that the LO Lattice averaged estimate\,\cite{WP25}:
\beq
a_\mu\vert^{hlbl}_{lo}\vert_{lattice}= (122.5\pm 9.0)\times 10^{-11},
\label{eq:hlbl-lattice}
\eeq
is slightly higher than the LO analytic one. 
\vspace*{-1.5cm}
\section{Total sum of  the hadronic contributions and SM predictions of $a_\mu$}
Adding separately the LO analytical and Lattice estimates to which we add the NLO and NNLO analytic estimates,
we obtain the total sum of the hadronic contributions $a_\mu^h$\,:
\vspace*{-0.5cm}
\begin{table}[H]
 \begin{center}
 \setlength{\tabcolsep}{2pc}
\begin{tabular}{lll}
&\\
\hline
\purple This work 2025& \blueda Mean Lattice&\blueda BMW20\\
\hline
\purple$7112\pm 38$& \blueda$7160\pm62$&  \blueda $7113\pm 56$ \\
\hline
\end{tabular} 
\end{center}
\label{tab:amu-lo}
\end{table}
\vspace*{-0.5cm}
There are a perfect agreement between our and the BMW results while the Lattice mean central value is higher by $48\times 10^{-11}$ but with a large error. The agreement with BMW si due to the fact that the higher value of $a_\mu\vert^{hvp}_{lo}$  is compensated by the lower value of $a_\mu\vert^{hlbl}_{lo}$ from an analytical analysis while in both cases of analytical and lattice calculations, we add  the same analytical NLO and NNLO contributions.
\section{Some other Standard Model Predictions to the Muon Anomaly}
\subsection*{\b The QED  contributions to  $a_\mu$}
\begin{figure}[hbt]
\begin{center}
\hspace*{-0.3cm}\includegraphics[width=2.9cm]{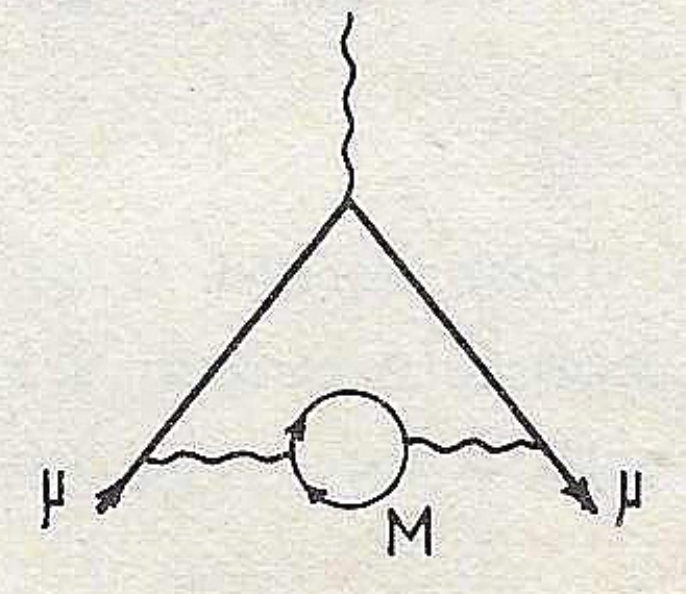}
\includegraphics[width=2.3cm]{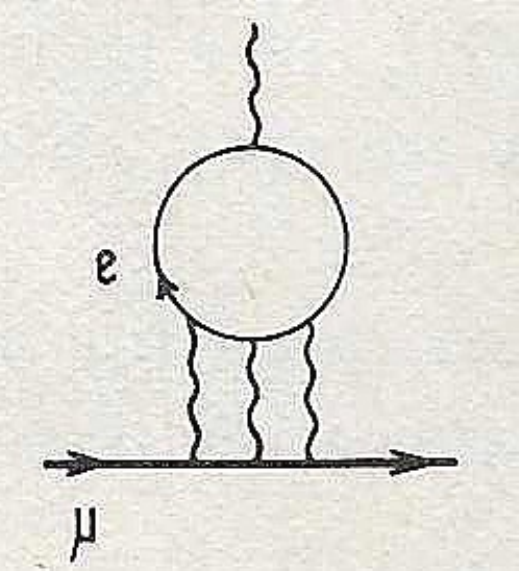}
\includegraphics[width=2.5cm]{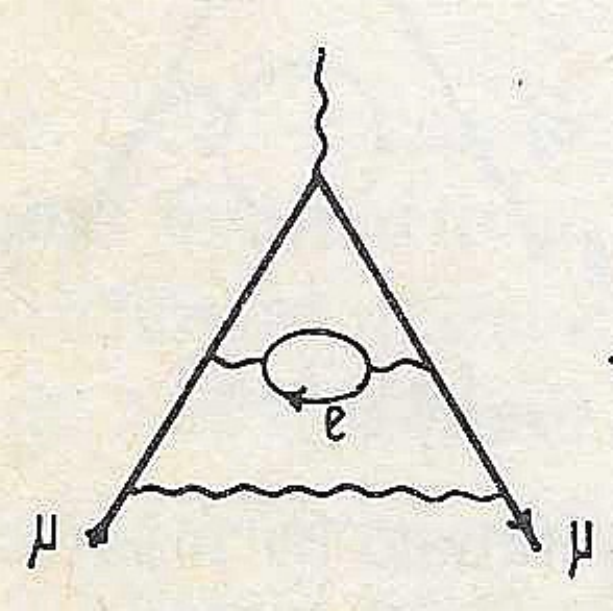}
\includegraphics[width=2.6cm]{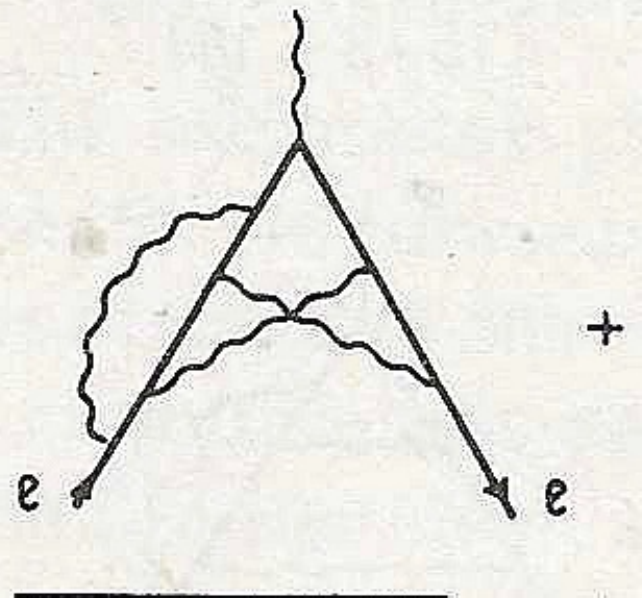}
\includegraphics[width=3cm]{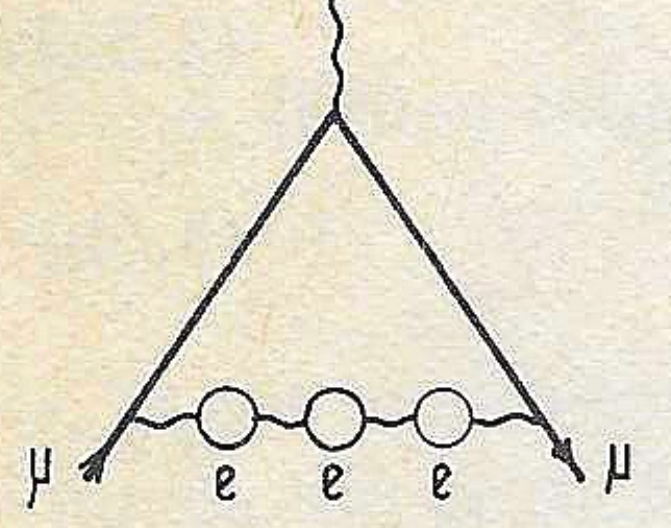}
\includegraphics[width=1.9cm]{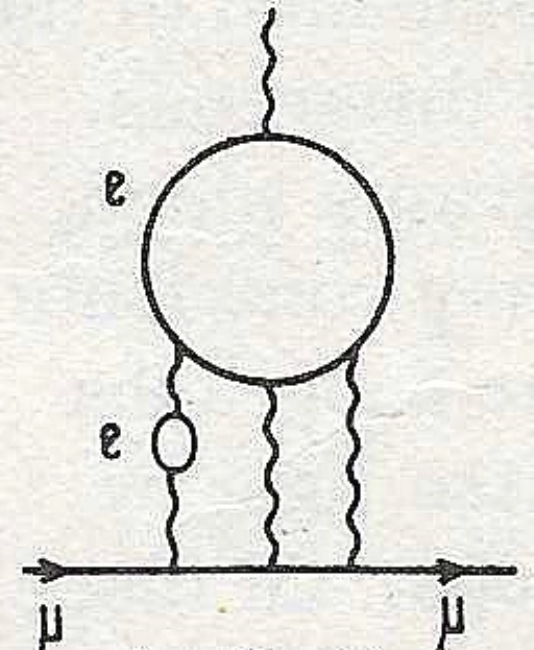} \boldmath+$\cdots$
\caption{\scriptsize NLO and N2LO  QED quantum corrections to $a_\mu$.}
\label{fig:amu-ho}
\end{center}
\end{figure}

A large amount of efforts have been done for calculating Higher Order (HO) corrections up 10th order $\alpha^5$ and on the estimate of the 12th order corrections (for reviews, see e.g.\,\cite{CALMET,DERAFAEL,AOYAMA,WP25}) (see e.g. some samples of Next-to-Leading (NLO) and Next-to-Next-to-Leading (N2LO) diagrams in Fig.\,\ref{fig:amu-ho}.
. 

The uncertainty on the measurement of $\alpha$ induces an uncertainty  $\Delta a_\mu \approx 0.137\times 10^{-11}$ which  is about the sie of the estimate of the 12th order term contribution. This feature leads Ref.\,\cite{WP25} to take the conservative value\,:
\beq
a_\mu^{qed}= (116 584 718.8\pm 2.0) \times 10^{-11}.
\eeq
\subsection*{\b The Electroweak (EW)  corrections to  $a_\mu$}
\begin{figure}[hbt]
\begin{center}
\includegraphics[width=9cm]{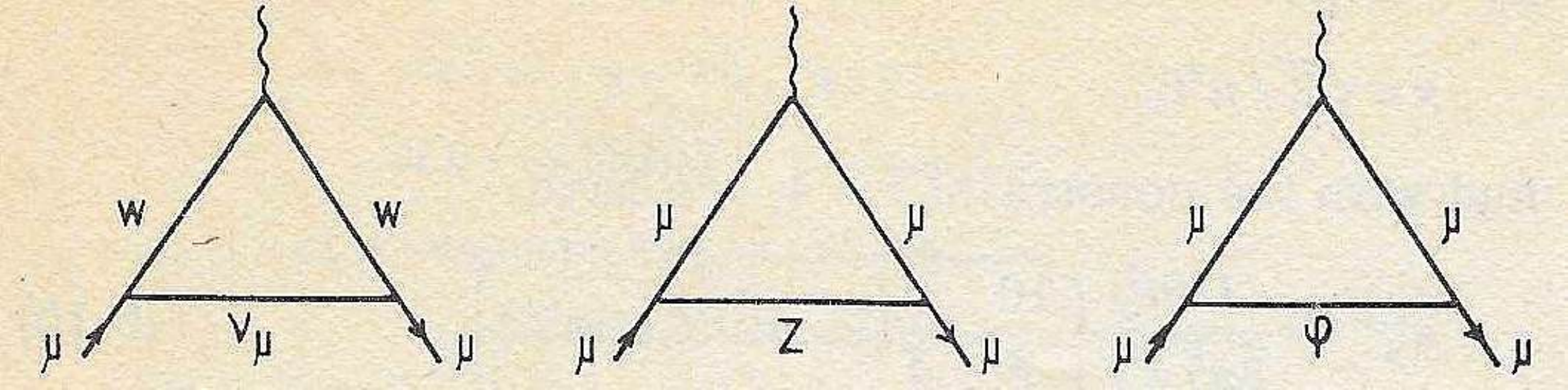}
\caption{\scriptsize LO electroweak corrections to $a_\mu$ with $W,Z$ and Higgs boson ($\phi$) exchanges.}\label{fig:amu-ew}
\end{center}
\vspace*{-0.5cm}
\end{figure}

These corrections are illustrated to Lowest Order (LO)  by the diagrams in Fig.\,\ref{fig:amu-ew}
Up to two virtual fermion and boson loop corrections and including the leading-log three-loop corrections, one obtains\,\cite{WP25,AOYAMA}:
\beq
a_\mu^{ew}=  (154.4\pm 0.4) \times 10^{-11}.
\eeq
\subsection*{\b Sum of the QED and Electroweak (EW)  corrections to  $a_\mu$}
From the previous analysis, we deduce the sum\,:
\beq
a_\mu^{qed}+a_\mu^{ew}= (16 584 873.2\pm 4.5) \times 10^{-11},
\eeq
which is 3.3 more precise than the recent experimental value in Eq.\,\ref{eq:final} but differs by:
\beq
\la a_\mu^{exp}\ra- a_\mu^{qed}-a_\mu^{ew} = (7198.3\pm 14.5_{exp}\pm 4.5_{th})\times 10^{-11}.
\label{eq:amuth}
\eeq
This discrepancy beween Experiment with the QED $\oplus$ EW contributions is expected to be compensated by the Hadronic contributions to $a_\mu$ and by some eventual ones beyond the Standard Model (BSM). 
\section{Attempt to give the  Standard Model (SM)  prediction of $a_\mu$}

We compare  in Fig.\,12 some different theoretical estimates versus experiment.  From left to right\,: Our Work, CMD3\,\cite{CMD3}, LATTICE (mean value from Ref.\cite{WP25}), BMW\,\cite{BMW}, TAU\,\cite{MASJUAN}, JEG\,\cite{JEGER26}, 
BABAR\,\cite{BABAR2}, KLOE, BESIII, SND quoted in \,\cite{WP25}, KNTZ\,\cite{NOMURA}, DHZ\,\cite{DAVIER2}, ALL (attempt to fit all $e^+e^-$ data)\,\cite{BZ26}. 
\begin{figure}[hbt]
\begin{center}
\hspace*{0.cm}\includegraphics[width=12cm]{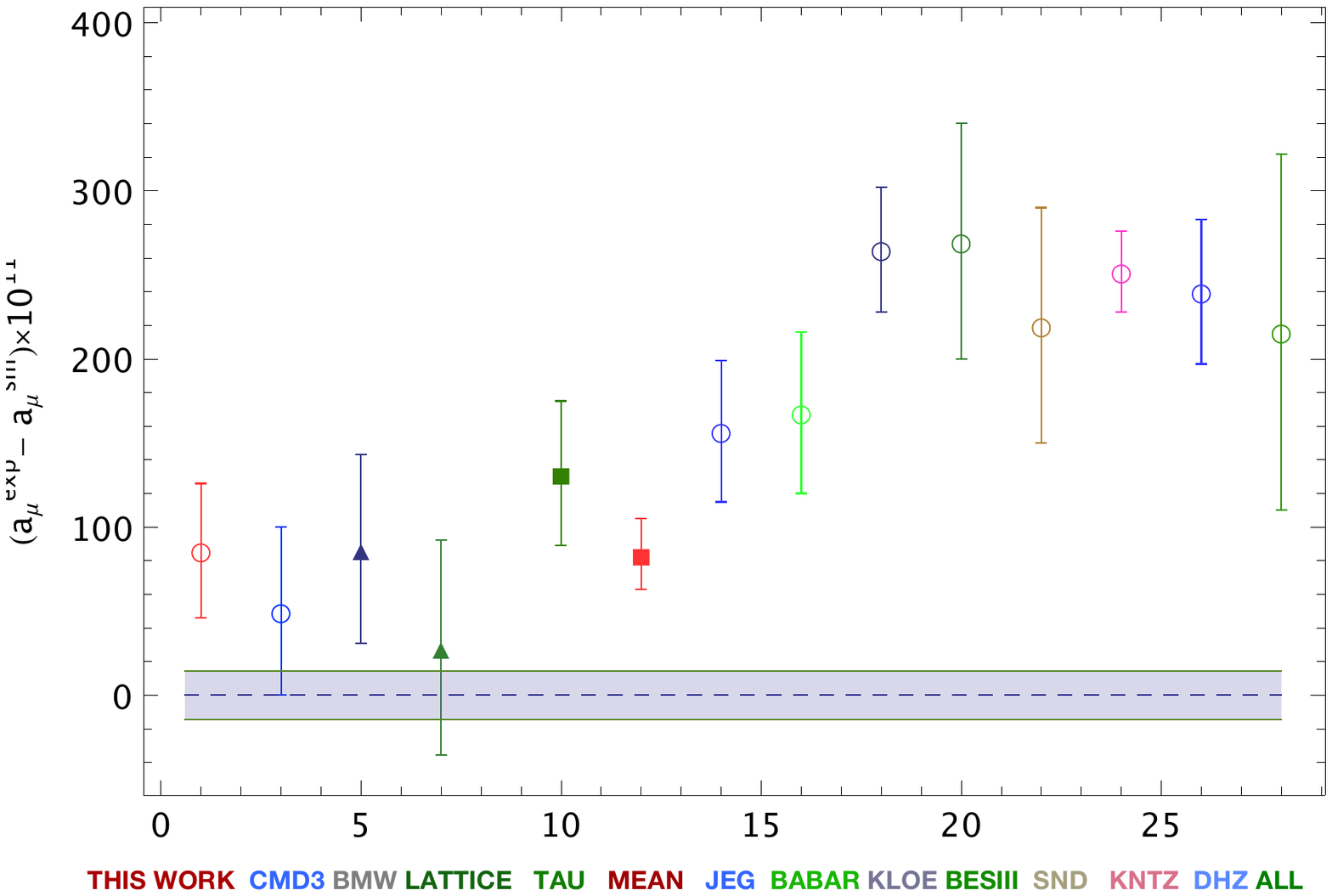}
\vspace*{-0.5cm}
\caption{Some theoretical estimates of $a_\mu$ using $e^+e^-$ and $\tau$-decay data and lattice calculations versus experiment}
\end{center}
\label{fig:amuth}
\end{figure} 
 
We show in Table\,3 some selected estimates which agree within the errors with our predictions. We can conclude that our analytical estimate based on $e^+e^-\to$ Hadrons data indicates a 2.1 $\sigma$ standard deviation from the SM prediction while the average of the ones quoted in Table 3 indicate a deviation of about  3.2 $\sigma$.  Adding to Table 3 the attempt to estimate all (inconsistent) $e^+e^-$ data\,\cite{BZ26}, the average becomes\,: 
\beq
\Delta a_\mu^{sm}=(90\pm21_{th}\pm 14.5_{exp})\times 10^{-11}.
\eeq
\vspace*{-1cm}
\begin{table} [hbt]
\label{tab:amu2}
 \setlength{\tabcolsep}{0.9pc}
   \begin{center}
\begin{tabular}{lclll}
&\\
\hline
Sources &$a_\mu\times 10^{11}$& $\Delta a_\mu^{sm}\equiv (a^{exp}_\mu-a^{sm}_\mu)\times 10^{11}$ \\
\hline
\oliva $FNAL$\,\cite{FNAL1}&\oliva $116592071.5\pm 14.5$& \\
\purple $This\, Work\,:\,e^+e^-\oplus Analytic_{\leq nnlo}$&\purple $116 591985.5\pm 38.2$&
\purple $ 86\pm 38_{th}\pm 14.5_{exp}$ \\
\blueli $CMD3$\,\cite{CMD3}&--& $ 50\pm 50_{th}\pm 14.5_{exp}$\\
 $Precise\, Lattice_{lo}$\,\cite{BMW} $\oplus Analytic_{\leq nnlo}$&$116 591984.5\pm 56$& $ 87\pm 56_{th}\pm 14.5_{exp}$ \\
 $Mean\, Lattice_{lo}$\,\cite{WP25} $\oplus Analytic_{\leq nnlo}$& $116 592033.5\pm 62$& $ 38\pm 62_{th}\pm 14.5_{exp}$ \\
  \oliva $Mean\, \tau$\,\cite{MASJUAN} $\oplus Analytic_{\leq nnlo}$&\oliva $116 591939.5\pm 42$&\oliva $ 132\pm 42_{th}\pm 14.5_{exp}$ \\
 \hline
 \purple$Average$ 
 &\purple $116 591987.7\pm 23.9$   &\purple $84.9\pm 21.3_{th}\pm 14.5_{exp}$ \\
\hline
\end{tabular}
\caption{Some recent estimates of the muon anomaly versus experiment consistent within the errors with our result.  }
\end{center}
\end{table}
\vspace*{-0.75cm}

\section{The LO Hadronic Vacuum Polarization contribution to the $\tau$  anomaly  $a_\tau\vert^{hvp}_{lo}$}
We extend the previous analysis to the $\tau$ anomaly. Due to the larger value of $M_\tau$ compared with $m_\mu$, the relative weight
of the low energy region in the integral (Eq.\,\ref{eq:amu-form}) decreases.  We have not redone the analysis in Ref.\,\cite{SN5} but only compile them  in Table\,\ref{tab:amu1} as the effect of the new fit of the data is almost negligible for the case of the muon.  Adding the estimate of the $P\gamma$ process of $(21\pm 0.4)\times 10^{-9}$ from\,\cite{NOMURA}, we deduce
\beq
a_\tau\vert^{hvp}_{lo}=(3516\pm 25)\times 10^{-9}.
\eeq
 It is remarkable to compare this result with the 1st estimate\,\cite{SN78}:
\beq
a_\tau\vert^{hvp}_{lo}=(3700\pm 400)\times 10^{-9},
\eeq
where the accuracy has increased by a factor 16 thanks to the experimental effort for measuring the cross-section of the $e^+e^-\to$ Hadrons process. 
\section{The LO Hadronic Vacuum Polarization contribution to $\Delta\alpha^{(5)}(M_Z^2)$}
This contribution is defined as:
\beq
\Delta \alpha_{had}^{(5)} (M_Z^2)=-\ga \frac{\alpha}{3\pi}\dr M^2_Z\int_{4m_\pi^2}^\infty \frac{R_{ee}(t)}{t(t-M_Z^2)},
\eeq
where $R_{ee}$ is the ratio of the $e^+e^-\to$ Hadrons over the  $e^+e^-\to \mu^+\mu^-$ total cross-sections and the upper index (5) corresponds to 5 flavours.
The contributions from different regions are shown in Table\,\ref{tab:amu1} where at the  $Z_0$ pole, we take the principal value of the integral\,:
\beq
\int_{4m_\pi^2}^{(M_Z- \Gamma_Z/2)^2}\hspace*{-1cm} dt\,f(t)+\int_{(M_Z+ \Gamma_Z/2)^2}^{\infty}\hspace*{-1cm}dt\,f(t),
\eeq
where $\Gamma =2.5$ GeV is the total hadronic $Z$-width. We add to the QCD continuum contribution the one of the $Z$-pole estimated to be\,\cite{YND17}:
\beq
\Delta \alpha_{had}^{(5)}\vert_{M_Z}= 29.2\times 10^{-5}.
\eeq
Then, we obtain the total sum given in Table\,\ref{tab:amu1}, to which, we add the $P\gamma$ contribution of $(4.4\pm 0.1)\times 10^{-5}$:
\beq
\Delta \alpha_{had}^{(5)} (M_Z^2)= (2770.7\pm 4.5)\times 10^{-5},
\eeq 
which improves and confirms our previous determination in Ref.\,\cite{SNalfa} (for a review, see e.g. Ref.\,\cite{JEGER}). 
\section{Summary and Conclusions}
\b We have reviewed the present status of the theoretical estimate of the muon and tau anomalies within the Standard Model by emphasizing the estimate of the hadronic contributions which are the main source of errors. We have completed the analysis by the re-estimate of the hadronic contributions to $\Delta \alpha^{(5)}_{hadrons}(M_Z^2)$

\b We notice that our results in Ref.\,\cite{SN5} are almost unaffected by the new fits of the precise\,\cite{BABAR} and new data from Belle 2\,\cite{BELLE} on $3\pi$ used to replace the NWA for estimating the $\omega$-meson. However, one has to add the contribution from $\pi^0,\eta \gamma$ not included in the CMD-3 data\,\cite{IGNATOV} and not taken into account in Ref.\,\cite{SN5}. 

\b We found that, within the recent precision data of $a_\mu$ from FNAL and our estimate of the LO Hadronic Vacuum Polarization (HVP) contributions from $e^+e^-\to$ hadrons,  a deviation from the Standard Model (SM) prediction by $2.1\sigma$, while an evarage of different determinations in Table\,3 gives a deviation of 2.8$\sigma$.

\b To have a sharper conclusion, one still needs to improve the measurement of the $e^+e^-\to$ Hadrons  data  in the  regions below the $\rho$-meson mass and around the $\rho-\omega$ mixing which are the main source of errors (see Table\,\ref{tab:amu1}), while the lattice calculations need to be significantly improved.  

\b The MUonE experiment project which aims to measure directly $a_\mu\vert^{hvp}_{l.o}$  and J-PARC (E-34) which plans to provide a new independent measurement of $a_\mu$ are  eagerly awaited for testing the theoretical estimates.


\end{document}